\pdfoutput=1
\documentclass[twocolumn]{aastex631}

\usepackage{amsmath}
\usepackage{amssymb}
\usepackage{hyperref}
\usepackage{color}
\usepackage{lmodern}
\usepackage{graphicx}

\newcommand{\code}[1]{\texttt{#1}}

\begin{document}


\title{Thermal Effects in Binary Neutron Star Mergers}


\author[0000-0001-5705-1712]{Jacob \surname{Fields}}
\email{jmf6719@psu.edu}
\affiliation{Department of Physics, The Pennsylvania State
University, University Park, PA 16802}
\affiliation{Institute for Gravitation \& the Cosmos, The
Pennsylvania State University, University Park, PA 16802}

\author[0000-0001-6734-3137]{Aviral \surname{Prakash}}
\affiliation{Department of Physics, The Pennsylvania State
University, University Park, PA 16802}
\affiliation{Institute for Gravitation \& the Cosmos, The
Pennsylvania State University, University Park, PA 16802}

\author[0000-0002-3327-3676]{Matteo \surname{Breschi}}
\affiliation{Theoretisch-Physikalisches Institut, Friedrich-Schiller-Universit{\"a}t Jena, 07743, Jena, Germany}
\affiliation{Scuola Internazionale Superiore di Studi Avanzati (SISSA), 34136 Trieste, Italy}
\affiliation{Istituto Nazionale di Fisica Nucleare (INFN), Sezione di Trieste, 34127 Trieste, Italy}

\author[0000-0001-6982-1008]{David \surname{Radice}}
\thanks{Alfred P.~Sloan fellow}
\affiliation{Institute for Gravitation \& the Cosmos, The
Pennsylvania State University, University Park, PA 16802}
\affiliation{Department of Physics, The Pennsylvania State
University, University Park, PA 16802}
\affiliation{Department of Astronomy \& Astrophysics, The
Pennsylvania State University,University Park, PA 16802}

\author[0000-0002-2334-0935]{Sebastiano Bernuzzi}
\affiliation{Theoretisch-Physikalisches Institut, Friedrich-Schiller-Universit{\"a}t Jena, 07743, Jena, Germany}

\author[0000-0003-0849-7691]{Andr\'e da Silva Schneider}
\affiliation{Departamento de Física, Universidade Federal de Santa Catarina, Florianópolis, SC 88040-900, Brazil}

\date{June 30, 2023}

\begin{abstract}
We study the impact of finite-temperature effects in numerical-relativity
simulations of binary neutron star mergers with 
microphysical equations of state and neutrino transport in which we vary the
effective nucleon masses in a controlled way. We find that, as the specific
heat is increased, the merger remnants become colder and more compact due to
the reduced thermal pressure support. Using a full Bayesian analysis, we
demonstrate that this effect will be measurable in the postmerger gravitational
wave signal with next-generation observatories  at signal-to-noise ratios of 15.
\end{abstract}




\section{Introduction.}
The extreme conditions found in neutron stars make them an ideal means for probing the nuclear equation
of state (EOS). Electromagnetic (EM) observations of pulsars have provided valuable information about
the mass distribution of neutron stars \citep{Lattimer.3.01,Alsing.4.18}, and recent results from NICER
offer constraints on their radii \citep{Riley.9.21,Ludlam.3.22,Salmi.9.22}. Binary neutron star (BNS)
mergers give additional astronomical constraints; the gravitational waves (GW) and EM counterpart of
GW170817 contained details about the EOS via tidal deformability measurements and ejecta
characteristics \citep{LIGO.10.17a,Margalit.10.17,Radice.11.17,De.4.18,LIGO.5.18,Dietrich.2.20,Kashyap.11.21}.

Due to the high Fermi temperature of matter in a neutron star,
constraints obtained from pulsars and BNS inspirals are informative of the zero-temperature equation of
state (EOS). On the other hand, temperatures as high as $100~{\rm MeV}$ might be reached in the
post-merger phase \citep{Perego.8.19}, making it a possible probe of the finite-temperature EOS.
Current GW detectors have not yet observed a BNS post-merger \citep{LIGO.10.17a,LIGO.10.17b,LIGO.3.20}.
Nevertheless, future detectors, like the proposed Einstein Telescope (ET) \citep{Maggiore.3.2020} and
Cosmic Explorer (CE) \citep{LIGO.7.16} detectors, will feature improved sensitivity at the higher
frequencies necessary to detect BNS post-merger signals. Inference on the EOS using post-merger data is
possible at (post-merger) signal-to-noise ratios (SNR) as low as 8 \citep{Breschi.5.22}. Additionally,
sensitivity upgrades to current instruments promise higher BNS detection counts with better sky
localization \citep{LIGO.living.18}.

State-of-the-art BNS merger simulations typically incorporate thermal effects via full
finite-temperature EOSs, often in the form of a table \citep{Bauswein.6.10,Sekiguchi.5.11}, and
realistic neutrino transport, such as via elaborate moment approximations
\citep{Foucart.7.16,Radice.3.22} or Monte Carlo methods \citep{Foucart.10.22}.
Many studies perform simulations with multiple finite-temperature EOSs to demonstrate sensitivity
(or lack thereof, as the case may be) of BNS merger observables under different scenarios, but the
different cold-temperature behavior of each EOS makes it difficult to attribute specific
outcomes to finite-temperature behavior \citep{Neilsen.5.14,Radice.11.17,Most.7.18,Perego.8.19,Hammond.8.21,Camilletti.4.22}.
Some studies have explored systematic changes in thermal effects through a so-called ``hybrid EOS'',
which extends a cold nuclear EOS to finite temperatures using an ideal gas component with a fixed
adiabatic constant $\Gamma_\text{th}$ \citep{Bauswein.6.10,Figura.5.20}. However, this is only a very
rough approximation, as the effective $\Gamma_\text{th}$ of a full finite-temperature EOS varies
considerably with density, temperature, and composition \citep{Carbone.8.19}.
\citet{Raithel.9.21} more recently considered
finite-temperature effects through a more sophisticated hybrid EOS based on an approximation to the
effective mass, but they do not fully explore how their model affects post-merger GW signals, nor is
this approach an entirely self-consistent model.
Furthermore, none of these studies incorporate all the relevant physics for modeling thermal effects,
particularly consistent neutrino transport.

In this \textit{Letter}, we present a first GR neutrino-radiation hydrodynamics study of
finite-temperature effects of a realistic nuclear EOS on BNS mergers through modifications to the
specific heat capacity. Our simulations show that an increased heat capacity results in denser, cooler
remnants. This leaves clear imprints on the GW signal in the post-merger phase, which we show can be
recovered in a parameter estimation pipeline tuned to a next-generation GW observatory.

\section{Methods.}
We select three non-relativistic Skyrme-type nucleonic EOSs built with the framework of
\citet{Schneider.11.19} and parameterized to produce the same cold nuclear matter bulk
properties but different specific heat content.
In Skyrme EOSs, the specific heat is controlled by the \textit{temperature-independent}
effective masses of neutrons and protons, $m^\ast_n$ and $m^\ast_p$, respectively.
These have a simple phenomenological description \citep{Constantinou.6.14, Schneider.11.19, margueron.2.18}
that depends only on two parameters and on the nucleonic number densities, $n_n$ and $n_p$ (or,
alternatively, the number density $n=n_n+n_p$ and the proton fraction $Y_p=n_p/n$ of matter), and
converge toward the vacuum nucleon masses at zero density.
The parameters were chosen to reproduce two nuclear matter observables at saturation density: the
effective mass for symmetric nuclear matter, $m^\ast=m_n^\ast\simeq m_p^\ast$, and the neutron-proton
effective mass splitting for pure neutron matter, $\Delta m^\ast$.
Guided by theoretical and experimental efforts \citep{margueron.2.18, Li.2.18, Huth.2.21, Zhang.6.21},
the selected EOSs probe the average and extreme, but still plausible, expected values for $m^\ast$, $m^\ast=\{0.55, 0.75, 0.95\}\,m_n$,
while fixing the yet poorly constrained $\Delta m^\ast$ to $0.10\,m_n$.
The same EOSs were used to study GW signals from core-collapse supernovae \citep{Eggenberger_Andersen.12.21}.

To first order, the baryonic contribution to the specific heat of degenerate matter
found in the core of a neutron star, which dominates over the lepton contribution, is
$c_v = \left(\frac{\pi}{3}\right)^{2/3}\frac{T}{n}\left(n_p^{1/3}m_p^\ast+n_n^{1/3}m_n^\ast\right)$,
see Eq.~(151) of \citet{Constantinou.6.14}.
Thus, all else being equal, increasing $m^\ast$ leads to a larger specific heat capacity for matter in
the merger remnant, to which we attribute the differences seen across our simulations.

Using the pseudospectral code \code{LORENE} \citep{Gourgoulhon.8.16}, we construct initial data
for equal-mass binary neutron star systems in quasicircular orbit with a gravitational mass of
$M=1.35$~M$_\odot$ per star. We evolve each binary using \code{THC\_M1} \citep{Radice.3.22,Zappa.10.22},
an extension of the \code{THC} numerical relativity code \citep{Radice.10.13,Radice.3.14} incorporating
neutrino transport via a moment formalism \citep{Thorne.2.81,Shibata.4.11}. The implementation in
\code{THC\_M1} makes use of the Minerbo closure for the radiation pressure tensor, which is exact in
the optically thick limit. Thus while our neutrino treatment is approximate overall, we can capture
effects such as neutrino trapping and dissipative effects from out-of-equilibrium weak reactions
in the BNS remnant exactly. Our runs, while not modeling
the magnetic field explicitly, also account for the effects of heating and angular momentum transport
from magnetohydrodynamic (MHD) turbulence via a general relativistic large eddy simulation (GRLES)
formalism calibrated with high-resolution GRMHD BNS simulations
\citep{Radice.3.17,Kiuchi.10.17,Radice.5.20}.

We perform each simulation
at two resolutions, designated as LR and SR, respectively corresponding to a grid spacing
of $\Delta x \approx 250$~m and $\Delta x \approx 180$~m in the finest refinement level, which covers
both stars during the inspiral and merger phases.
We also run identical SR simulations with our older M0 solver \citep{Radice.1.16} to validate
our results; though the solver is less accurate and does not properly capture effects such as neutrino
trapping, these runs support the major results of the M1 runs.

\section{Results.}
\begin{figure}
  \includegraphics[width=\columnwidth]{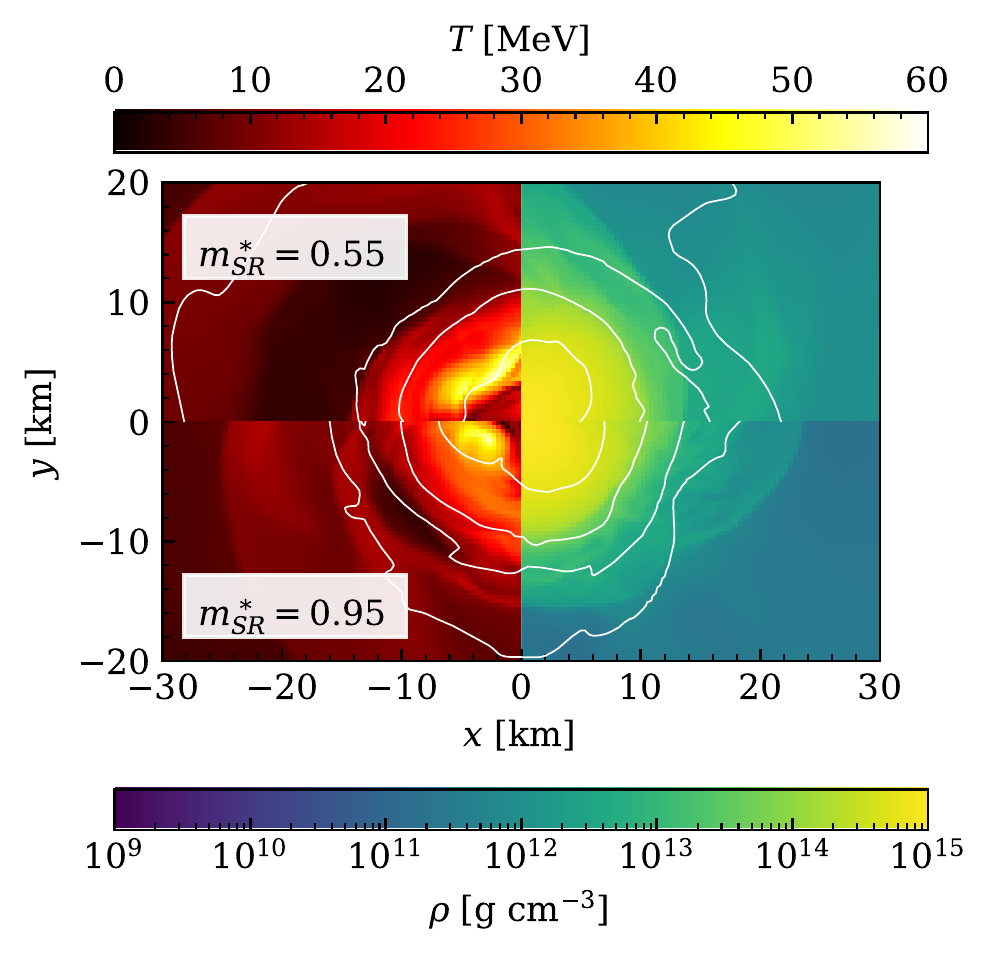}
  \includegraphics[width=\columnwidth]{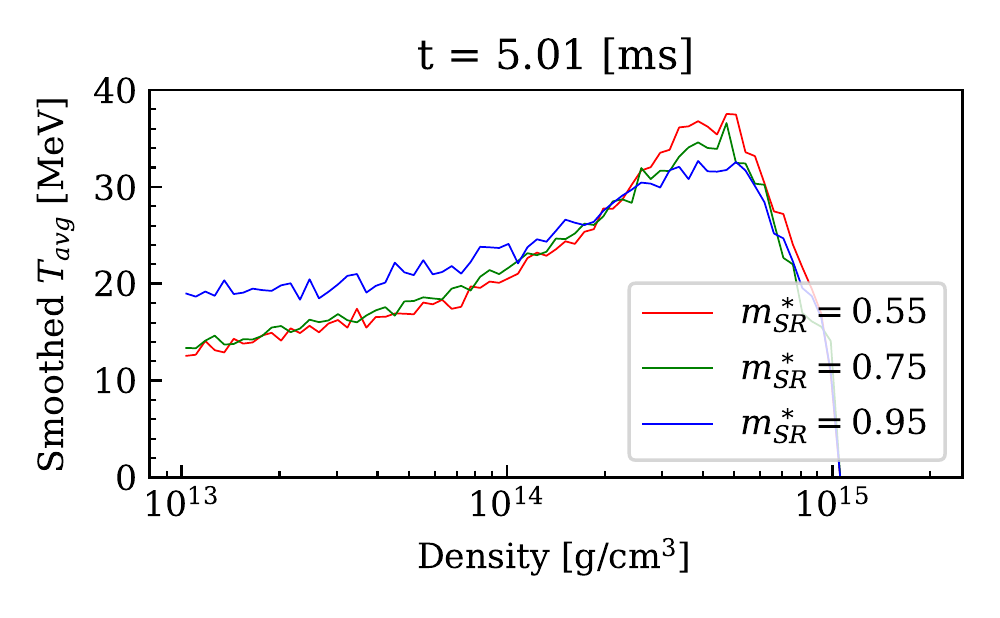}
  \caption{\label{fig:dtslice} Temperature and rest-mass density slices in the $x$-$y$ plane of the
           $m^\ast=0.55$ and $m^\ast=0.95$ SR
           simulations at approximately 5 ms post-merger (top) and average temperature as a function
           of density (bottom). Contour lines correspond to rest-mass
           densities $\rho=\{10^{12},10^{13},10^{14},5\times10^{14}\}\
           {\text{g}\ \text{cm}}^{-3}$. For visual clarity, the $m^\ast=0.75$ model is not shown in
           the slice plot.
           }
\end{figure}

\begin{figure}
  \includegraphics[width=\columnwidth]{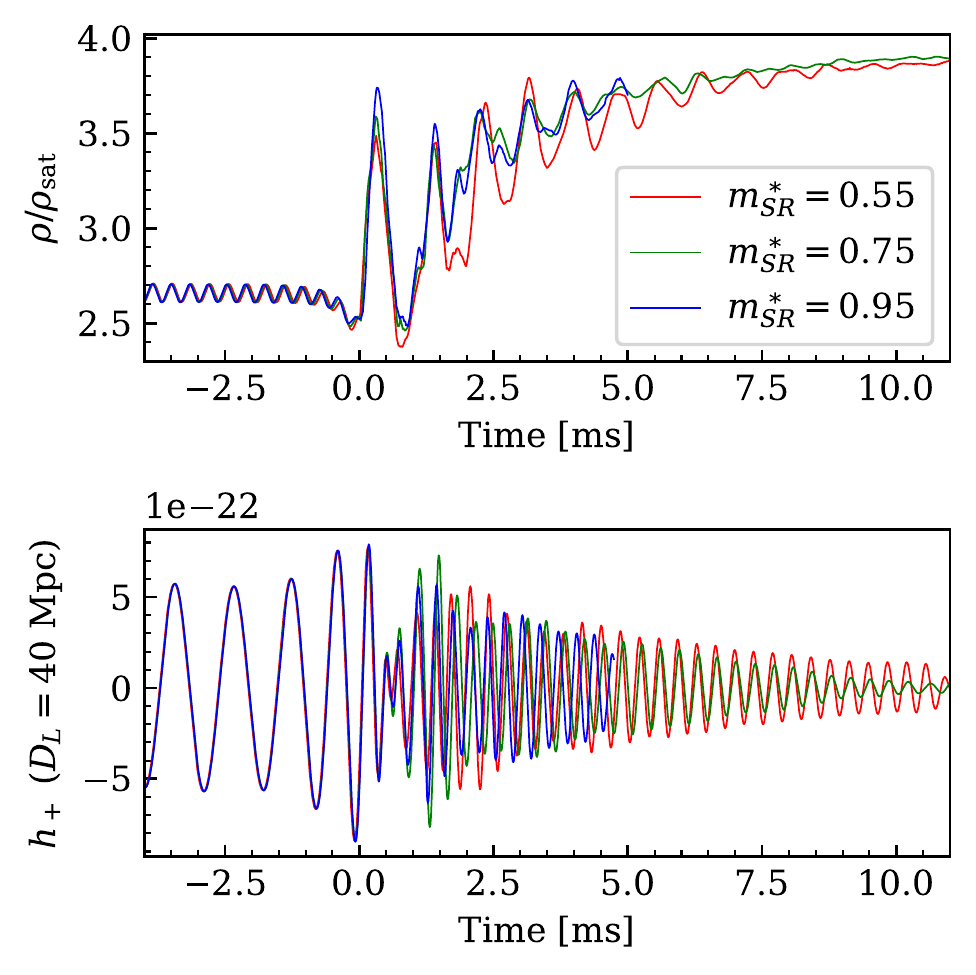}
  \caption{\label{fig:strainrhoeb} The maximum rest-mass density of the SR simulations (top) and the
           the gravitational-wave strain of the $\ell=2$, $m=2$ mode (top).}
\end{figure}

Figure~\ref{fig:dtslice} shows slices of the $x$-$y$ plane for the $m^\ast=0.55$ and $m^\ast=0.95$ SR
simulations at approximately $5$~ms post-merger and average temperatures for all three values of 
$m^\ast$\footnote{Unless otherwise noted, $m^\ast$ has units of $m_n$, the neutron mass.}.
Even at this relatively early post-merger time, the density of the remnant's inner core is noticeably
larger in the $m^\ast=0.95$ model. Conversely, the temperature is lower for the $m^\ast=0.95$ compared to
the $m^\ast=0.55$. The $m^\ast=0.75$ model has an intermediate behaviour compared to the other two.
Interestingly, we find that the temperature trend reverses at lower densities (Figure~\ref{fig:dtslice}),
possibly because of the larger compactness of the merger remnant in the larger $m^\ast$ models. 
In other words, for higher $m^\ast$, surface material (where the differences between EOS models are much smaller) falls
deeper into the gravitational potential, becoming hotter in the process. A similar
trend was reported for CCSN simulations \citep{Schneider.11.19}.
The top panel of Figure~\ref{fig:strainrhoeb} shows that the maximum rest-mass density is highest in the
$m^\ast=0.95$ model and lowest for $m^\ast=0.55$ at most post-merger times. As suggested in this plot,
all of the models, independent of resolution or neutrino treatment, produce long-lived remnants.

\begin{figure*}
  \centering
  \includegraphics[width=0.6\textwidth]{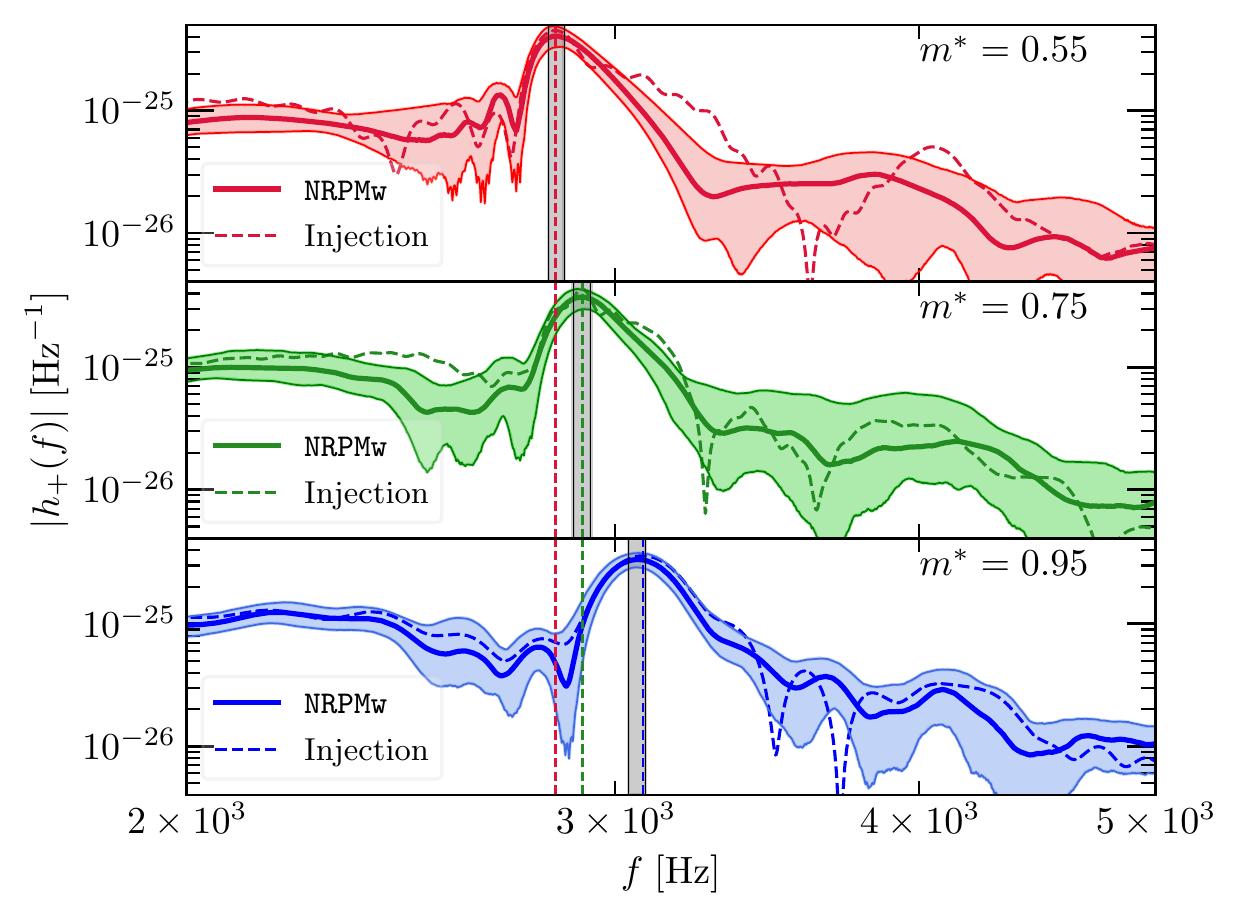}
  \caption{\label{fig:f2} The reconstructed GW spectrum of the $\ell=2$, $m=2$ mode using the $\tt{NRPMw}$ model. The colored solid lines represent the median waveform
  and the colored shaded regions represent the 90$\%$ credible intervals on the posterior distribution of the spectra computed from the recovered parameter space of $\tt{NRPMw}$. The colored dashed limes represent the injected spectra at an SNR of 15. Additionally, shown are the postmerger peak frequencies $f_2$ (in vertical dashed lines) and the 90$\%$ credible intervals (in grey) on the posterior distribution of $f_2$ from the reconstructed waveforms.}
\end{figure*}

The GW strain in Figure~\ref{fig:strainrhoeb} (bottom panel) demonstrates identical behavior in the
inspiral for all three models, but the waveforms begin to deviate in the post-merger due to
finite-temperature effects in the EOS. The differences in morphology include frequency evolution,
amplitude, damping times, and modulations. This is more quantitatively seen in the GW spectrum
(see Figure~\ref{fig:f2}), where there is a clear rightward shift in the peak postmerger frequency, $f_2$, as $m^\ast$
increases.
Table~\ref{tab:f2} contains $f_2$ for both the LR and SR simulations. Although the
precision of the NR waveform is limited by finite resolution and step size, the shift of
$\Delta f \gtrsim 10$~Hz is robust across resolutions, which suggests it is an effect of the EOS and
not an
artifact of the simulations. The M0 SR simulation demonstrates a similar trend, although the
$m^\ast=0.75$ run is much closer to the $m^\ast=0.95$ model than either of the M1 cases. This provides
an important sanity check of our results, but we stress that it is only qualitative; the M1 results
should be used for quantitative analysis of post-merger effects, as a self-consistent approach which
accounts for neutrino trapping noticeably changing the remnant's evolution (see \citet{Zappa.10.22};
the appendix also contains a more detailed comparison).

To detect these thermal effects in the postmerger via next-generation GW detectors and possibly constrain the $m^\ast$
nuclear parameter,
we perform full Bayesian inference on the postmerger GW signals. To compute injections, we extract the postmerger waveforms from the SR
simulations by applying a Tukey window to suppress the inspiral and spline interpolate the GW waveforms to a sampling rate of 16384 Hz. We further zero pad the
signals to a segment of 1 s. For brevity, we consider here only noise-less injections.
For parameter estimation, we employ the publicly available code \code{bajes} \citep{Breschi.9.21} and use the $\tt{UltraNest}$ \citep{johannes_buchner_2021_4636924}
sampler available as part of the $\tt{bajes}$ pipeline. We recover the injections by using the postmerger model $\tt{NRPMw}$ from
\citep{Breschi.5.22} and compute posteriors on its parameters. We inject all signals at a luminosity distance corresponding to a post-merger SNR of
15 (which corresponds to an inspiral SNR of $\mathcal{O}(100)$, see \citet{Maggiore.3.2020}), assuming the power spectral density of ET-D \citep{Hild:2010id} to simulate the detector response. The priors are set in accordance to \citep{Breschi:2022ens}, section-IIB. In Figure \ref{fig:f2}, we show the reconstructed waveforms from $\tt{NRPMw}$ evaluated on the parameter space of the recovered posterior samples.
We list the recovered SNRs and $f_2$ values in Table~\ref{tab:f2}. At $\text{SNR}=15$, the injected
spectrum's $f_2$ frequencies sit well within the $90\%$ credible intervals of the distribution of reconstructed waveform's $f_2$
frequencies. Furthermore, these intervals do not overlap, indicating that all three waveforms are clearly
distinguishable at a post-merger $\text{SNR}=15$ with $90\%$ credibility.
\begin{table}
\begin{ruledtabular}
  \caption{\label{tab:f2} Peak post-merger frequencies ($f_2$) of the gravitational wave spectrum for
           the LR (M1), M0, and SR (M1) simulations and the NRPMw model.
           For reference, the recovered matched-filter
           $\text{SNR}$ values and corresponding luminosity distance $D_L$ are also provided. For
           consistency, all peaks are measured after suppressing the inspiral.
           We also provide the mismatch, $\mathcal{M}$ (see \citet{Lindblom.9.08,Damour.1.11}), between
           the $f_2^\text{SR}$ runs, with one row measured against $m^\ast=0.75$ and the other against $m^\ast=0.95$.}
  \begin{tabular}{ll|lll}
    $m^\ast$ & ($m_n$) & $0.55$ & $0.75$ & $0.95$ \\
    \hline
    $f_2^\text{LR}$ & (kHz) & $2.862$ & $2.908$ & $2.921$ \\
    $f_2^\text{M0}$ & (kHz) & $2.864$ & $2.966$ & $2.974$ \\
    $f_2^\text{SR}$ & (kHz) & $2.835$ & $2.908$ & $3.080$ \\
    $\mathcal{M}^\text{SR}_{0.75}$ & & $0.04$ & N/A & $0.08$ \\
    $\mathcal{M}^\text{SR}_{0.95}$ & & $0.09$ & $0.08$ & N/A \\
    $f_2^\text{NRPMw}$ & (kHz) & $2.83^{+0.02}_{-0.02}$ & $2.91^{+0.02}_{-0.02}$ & $3.06^{+0.02}_{-0.02}$ \\
    $\text{SNR}_\text{mf}$ & & $13.6^{+1.6}_{-2.7}$ & $13.4^{+1.6}_{-2.5}$ & $13.8^{+1.8}_{-2.2}$ \\
    $D_L$ & (Mpc) & $56.353$ & $56.730$ & $49.655$
  \end{tabular}

\end{ruledtabular}
\end{table}

\section{Discussion.}
We have shown that $m^\ast$ significantly influences the GW signals in BNS mergers, and we have
analyzed this effect in detail for $f_2$. The relationship between $m^\ast$ and $f_2$ is easily
explained in terms of the specific heat. Increasing the specific heat appears to soften the equation of
state; because it requires more energy to increase the temperature, there is less thermal pressure
available to support the star, thus producing a more rapidly rotating and compact remnant with lower
temperatures, in agreement with core-collapse supernovae studies
\citep{Eggenberger_Andersen.12.21, Schneider.11.19, Yasin.3.20}.

The EOSs we use have a simple relationship between $m^\ast$ and the specific heat capacity which may
not be representative of the true nuclear EOS, which is expected to be both density and temperature
dependent \citep{Carbone.8.19}.
Nevertheless, this study serves as a proof of concept
demonstrating that future detectors like ET and CE can use $f_2$ to constrain the finite-temperature
EOS. We also reiterate that this effect \textit{only} affects the finite-temperature evolution, which
will not be observable until next-generation detectors sensitive to the post-merger phase come online.
Furthermore, we also note that this study does not provide a method for measuring $m^\ast$ from $f_2$;
we only assert that $m^\ast$ leaves an imprint on $f_2$.

We do acknowledge some limitations in our work; most notably, the length of the $m^\ast=0.95$ M1
simulation is much shorter than both the $m^\ast=0.55$ and $m^\ast=0.75$ runs due to a limitation in
the M1 neutrino solver which introduced unphysical effects past $5$~ms post-merger. This short length
may explain why $f_2^{m^\ast=0.95}$ increases between the LR and SR runs while both $f_2^{m^\ast=0.55}$
and $f_2^{m^\ast=0.75}$ instead decrease, as a shorter signal will introduce a extra spread of
frequencies to the power spectrum and possibly shift the peak.
Another possible limitation concerns the omission of magnetic fields. Our simulations include a GRLES model which can
account for some, but not all MHD effects. On the other hand, other simulations suggest that
MHD effects on the post-merger gravitational wave frequency are negligible for realistic initial
magnetic field strengths \citep{Palenzuela.7.22}.

We may consider several avenues for future work. Longer simulations would allow us to investigate
the ejecta and consider possible effects on EM counterparts, as well as possible thermal effects
on the lifetime of the remnants. Additionally, \citet{Zappa.10.22} indicate that
resolution has a prominent influence on post-merger evolution, including collapse time and disk
formation. Although we validate our results here with two resolutions, accurately determining the 
precise value of $f_2$ for each model would require higher resolution calculations. To investigate our 
hypothesis that this study's results are general, future simulations could also explore other EOS 
models with tunable finite-temperature behavior.

\section*{Note Added}
While this manuscript was under review, \citet{Raithel.6.23} announced new results on the impact of
thermal effects on $f_2$. Their work uses a semi-analytic prescription for the EOS and neglects
neutrinos, but it is in good agreement with our findings.

\begin{acknowledgments}
The authors thank Albino Perego, Ish Gupta, Arnab Dhani, Rahul Kashyap and Bangalore Sathyaprakash for helpful discussions.
DR acknowledges funding from the U.S. Department of Energy, Office of Science,
Division of Nuclear Physics under Award Number(s) DE-SC0021177 and from the
National Science Foundation under Grants No. PHY-2011725, PHY-2020275,
PHY-2116686, and AST-2108467.
SB acknowledges support by the EU H2020 under ERC Starting Grant,
no.~BinGraSp-714626 and from the Deutsche Forschungsgemeinschaft DFG project
MEMI number BE 6301/2-1.
MB acknowledges support by the PRO3 program ``DS4Astro''
of the Italian Ministry for Universities and Research.
Simulations were performed on PSC Bridges2 (NSF XSEDE allocation TG-PHY160025).
This research used resources of the National Energy Research Scientific
Computing Center, a DOE Office of Science User Facility supported by the Office
of Science of the U.S.~Department of Energy under Contract
No.~DE-AC02-05CH11231.  Computations for this research were performed on the
Pennsylvania State University’s Institute for Computational and Data Sciences’
Roar supercomputer.

\end{acknowledgments}

\appendix
\onecolumngrid
\section{Neutrino Effects}
The presence of a trapped neutrino gas in optically-thick regions can affect the temperature and
composition of the remnant \citep{Perego.8.19,Zappa.10.22}. Many BNS simulations today employ leakage
schemes, which do not explicitly model neutrino radiation, but instead estimate a cooling rate based on
the local optical depth \citep{Sekiguchi.7.11,Neilsen.5.14,Palenzuela.8.15,Radice.1.16,Radice.12.18}.
Because they do not perform consistent radiation transport, leakage schemes cannot capture the
trapped neutrino gas in the remnant or potential out-of-equilibrium effects. The more accurate M1 scheme,
however, directly models neutrino transport and can therefore capture both these effects
\citep{Radice.3.22, Zappa.10.22}. Because our study represents the first investigation of thermal effects
in BNS mergers with self-consistent neutrino transport, we will demonstrate here the influence of
neutrinos on our results by comparing our M0 (a leakage-style scheme) simulations to our M1 runs.

\begin{figure}
  \centering
  \includegraphics[width=0.9\columnwidth]{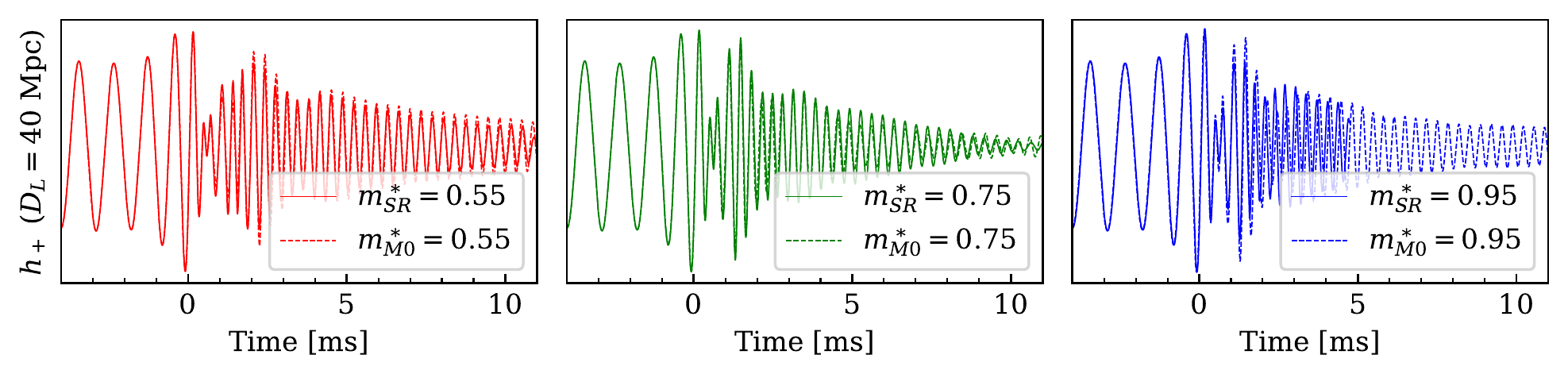}
  \caption{\label{fig:straincomparison} The GW strain of the $\ell = 2$, $m = 2$ mode at
           $D_L=40~\text{Mpc}$ for both the M1 (solid lines) and M0 (dashed lines) SR simulations.}
\end{figure}

In Fig.~\ref{fig:straincomparison}, we show the GW strain for both the M0 and M1 simulations. All three
values of $m^\ast$ show identical behavior in the inspiral, which is to be expected; neutrinos
emissions are highly sensitive to the temperature, which is quite low prior to merger. In the
post-merger signal, the $m^\ast=0.55$ and $m^\ast=0.75$ differ only very slightly in amplitude and
frequency. The $m^\ast=0.95$ runs show somewhat stronger deviations but are still quite similar in
overall morphology. Nevertheless, the shifts in $f_2$ (see Table~I in the main text) for
changes in the neutrino solver are within a factor of two or three of the shifts due to changing
$m^\ast$; were these results used to calibrate some model for determining $m^\ast$ from $f_2$, it may
result in a fairly significant error.

\begin{figure}
  \includegraphics[width=0.45\columnwidth]{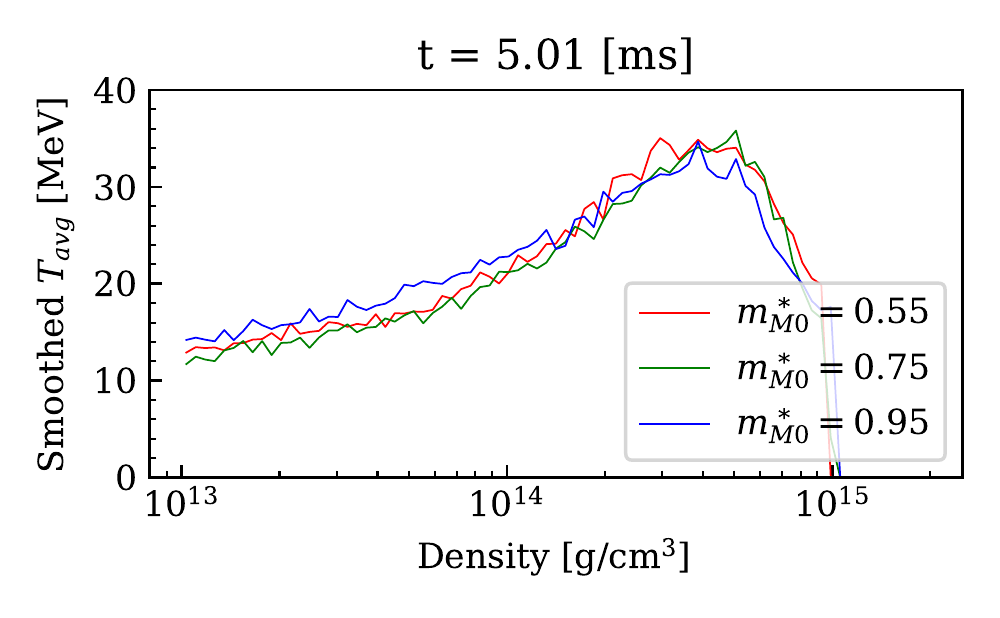}
  \hfill
  \includegraphics[width=0.45\columnwidth]{avg_temp_m1.pdf}
  \caption{\label{fig:tempcomparison} The average temperature as a function of density at
           $t\approx 5~\text{ms}$ post-merger for the M0 (left) and M1 (right) SR simulations. Each
           average was calculated by constructing two-dimensional histograms in temperature and
           density, then averaging over the temperature with a weight corresponding to the mass of
           each density-temperature bin. To reduce noise in the data, five equally-spaced timesteps
           (with $\Delta t\approx0.05~\text{ms}$) centered on $t\approx5~\text{ms}$ were averaged
           together.}
\end{figure}

In Fig.~\ref{fig:tempcomparison}, we show the average temperature as a
function of density at $5~\text{ms}$ post-merger for both the M0 and M1 SR simulations. We calculate 
this temperature by constructing two-dimensional histograms in temperature and density, then performing
a weighted average over the temperature for each density bin. This data is smoothed by averaging five
time steps centered around $5~\text{ms}$. The M1 data demonstrates a clear trend at higher densities 
($\rho \gtrsim \rho_\text{sat}$) where higher values of $m^\ast$ lead to lower temperatures, and there 
is limited evidence suggesting an inverted trend in the outer layers of the star 
($\rho \lesssim \rho_\text{sat}$), a result which is consistent with core-collapse supernovae 
simulations \citep{Schneider.11.19}. One possible explanation is due to the increased compactness for
higher $m^\ast$; material near the surface (where the density is low enough that $m^\ast$
has a much weaker effect on the EOS) falls deeper into the gravitational potential and heats up more as
$m^\ast$ increases. However, the high-density trend is
obscured in the M0 data, and there is no evidence of the low-density trend, as the $m^\ast=0.55$
temperature is often higher than the $m^\ast=0.75$ temperature at lower densities (where the M0 scheme
suppresses neutrino absorption).

\begin{figure}
  \includegraphics[width=0.45\columnwidth]{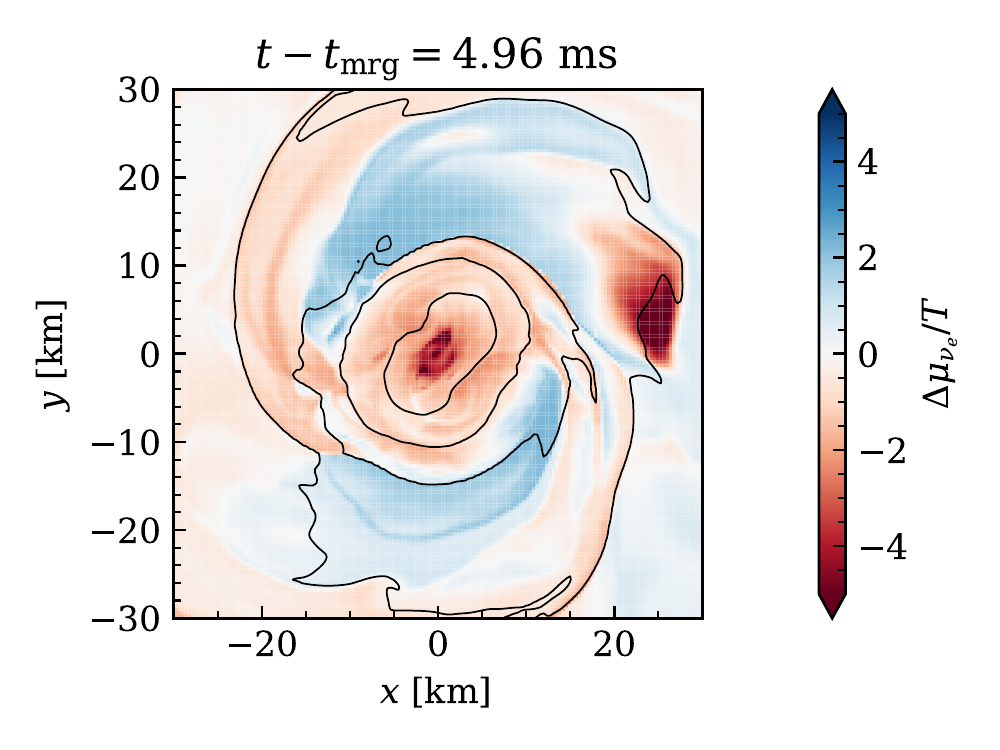}
  \hfill
  \includegraphics[width=0.45\columnwidth]{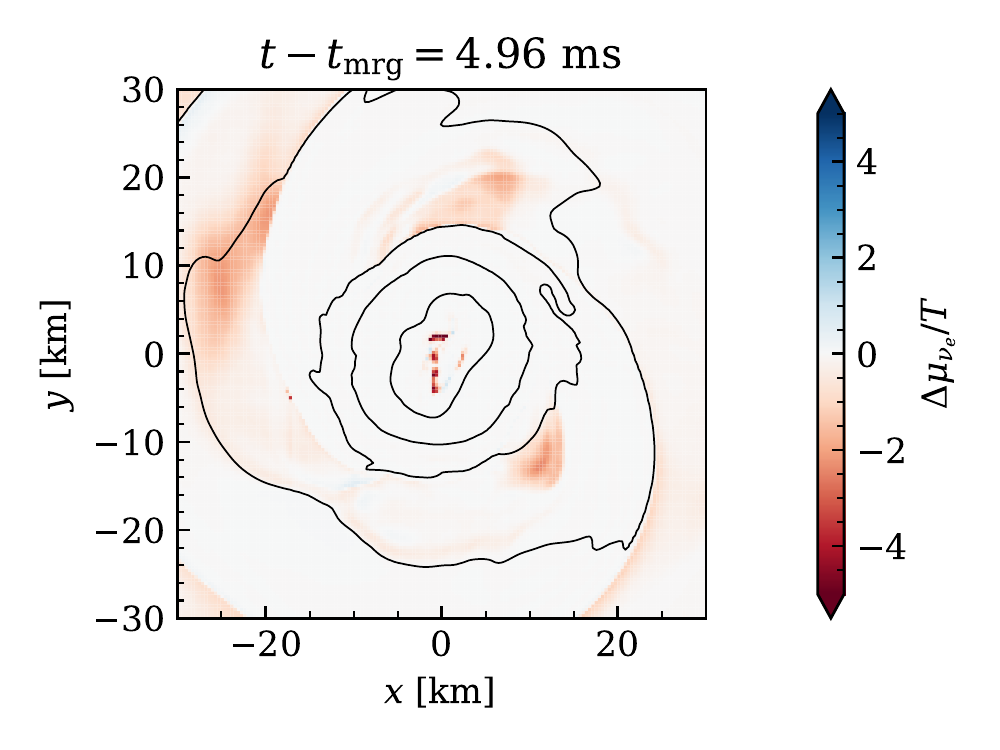}
  \includegraphics[width=0.45\columnwidth]{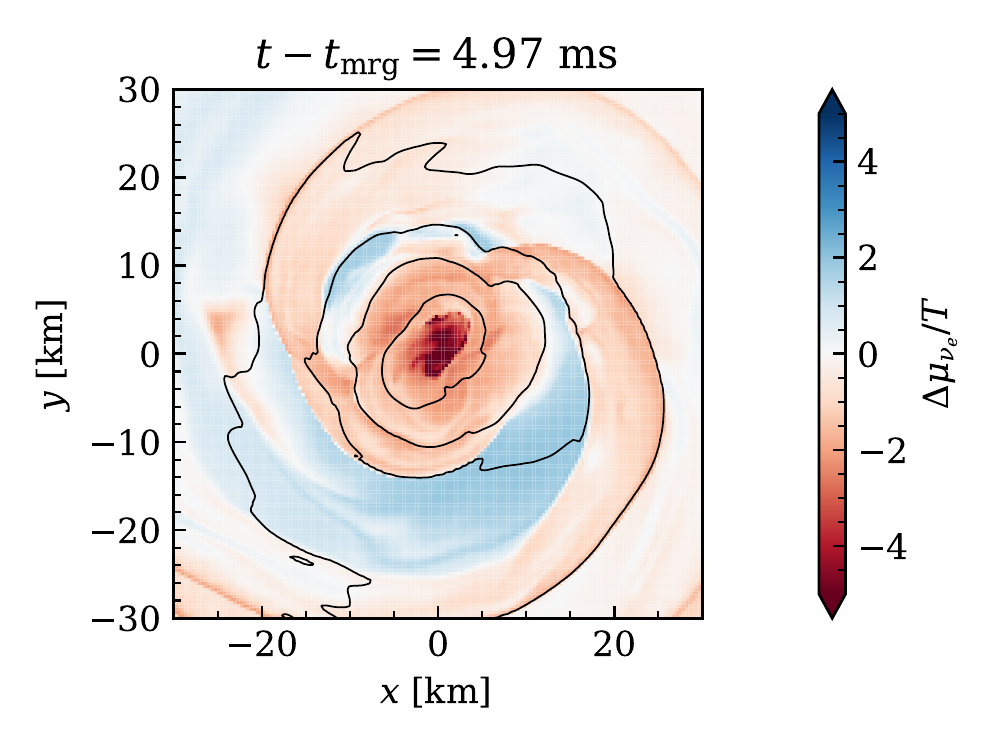}
  \hfill
  \includegraphics[width=0.45\columnwidth]{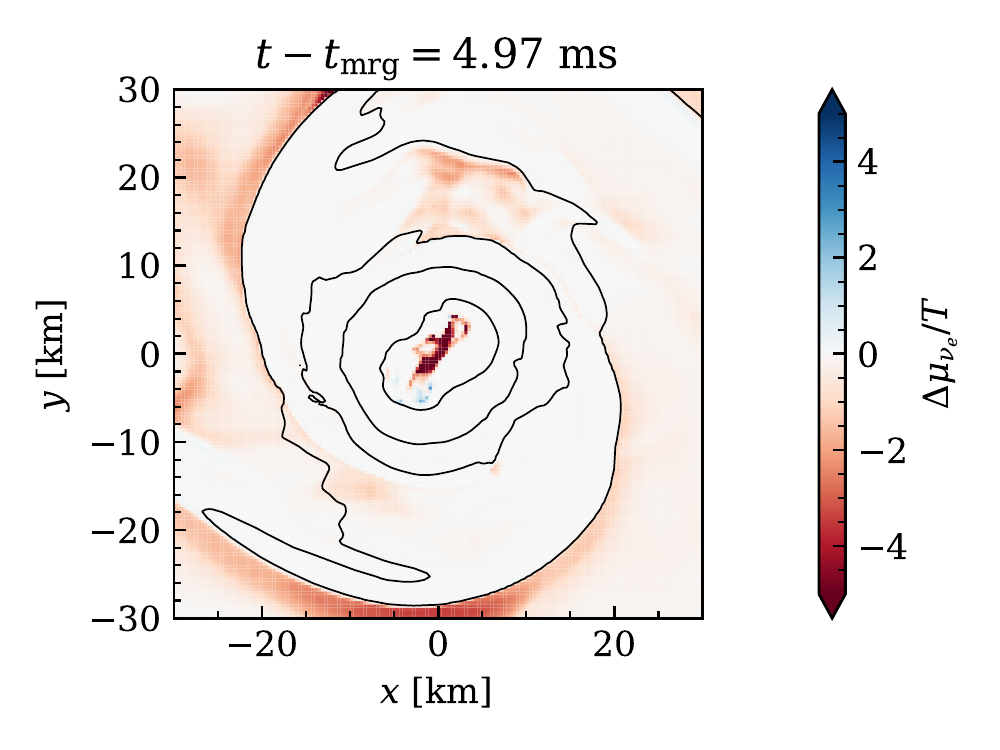}
  \includegraphics[width=0.45\columnwidth]{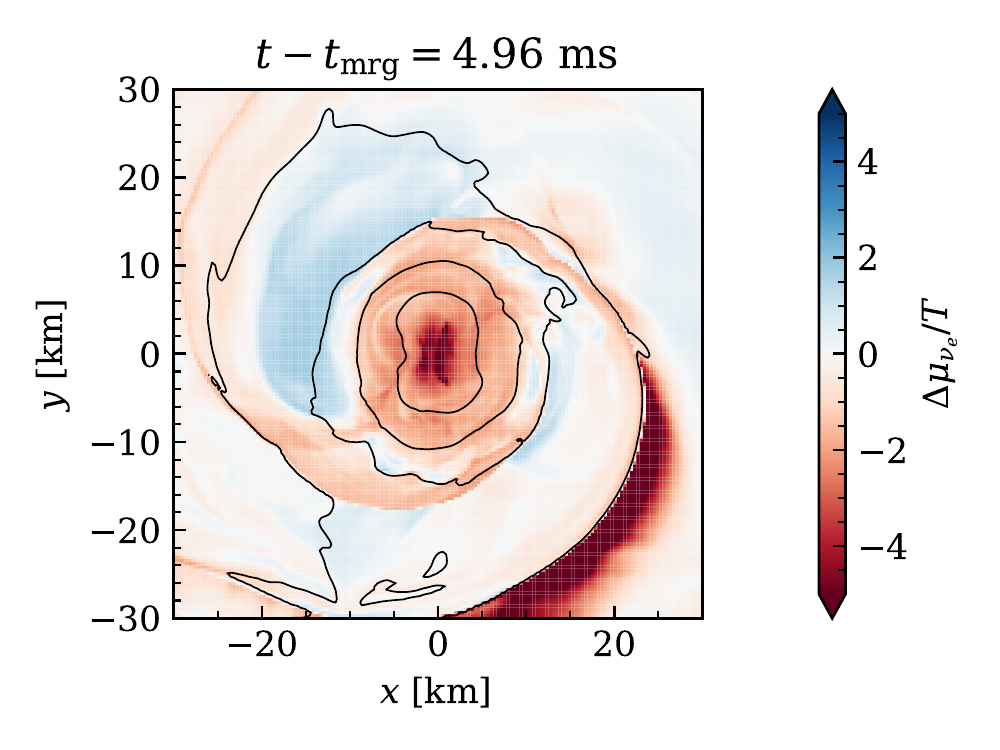}
  \hfill
  \includegraphics[width=0.45\columnwidth]{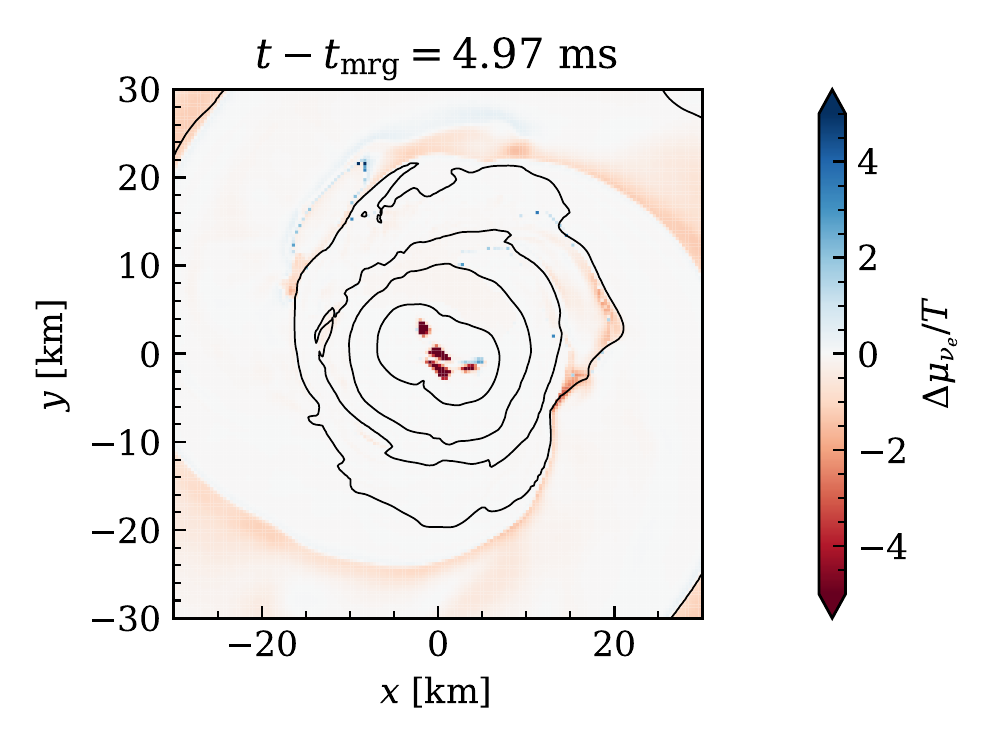}
  \caption{\label{fig:equilibrium} The deviation from weak equilibrium $\Delta \mu_{\nu_e}/T$ between
           M0 (left) and M1 (right) for $m^\ast=0.55$ (top), $m^\ast=0.75$ (middle), and $m^\ast=0.95$
           (bottom) at $t\approx 5~\text{ms}$. The contour lines correspond to the rest-mass densities
           $\rho=\{10^{12},10^{13},10^{14},5\times10^{14}\}\
           {\text{g}\ \text{cm}}^{-3}$.}
\end{figure}

In weak equilibrium
\begin{equation}
  \mu_{\nu_e} = \mu_p + \mu_e - \mu_n,
\end{equation}
where $\mu_i$ is the chemical potential for particle species $i$, with $p$, $e$, $n$, and $\nu_e$
respectively representing the proton, electron, neutron, and electron neutrino.
On the other hand, for a trapped
neutrino gas in thermal equilibrium with the nucleonic matter of density $\rho$, temperature $T$, and
electron fraction $Y_e$, we have
\begin{equation}
  \label{eq:thermeq}
  Y_{\nu_i} = \frac{4\pi m_b}{\rho} \left(\frac{k_\text{B}T}{hc}\right)^3
              F_2\left(\frac{\mu_{\nu_i}}{T}\right),
\end{equation}
where $m_b$ is the average baryon mass and $F_i(x)$ is the $i^\text{th}$ Fermi function.
This suggests that one can monitor the deviation from equilibrium by calculating the quantity
\begin{equation}
  \Delta \mu_{\nu_e} = \mu_p + \mu_e - \mu_n - \mu_{\nu_e}^T,
\end{equation}
where $\mu_{\nu_e}^T$ is $\mu_{\nu_e}$ calculated under the assumption of thermal equilibrium
(Eq.~\ref{eq:thermeq}) using the evolved neutrino fractions from M1. For M0 this quantity should be set
to zero, as the trapped neutrino gas is not explicitly modeled. We plot this deviation for $\nu_e$ in
Fig.~\ref{fig:equilibrium}. M0 shows noticeable deviations from weak equilibrium throughout the bulk of
the remnant. This shows that the M0 simulations are not correctly capturing the thermal equilibrium of
matter. On the other hand, the M1 simulations show that matter and neutrinos  are in equilibrium in most
regions in the remnant. We therefore conclude that a proper investigation of thermal effects must include
full neutrino transport.

\bibliography{references_apjl}

\end{document}